\documentclass{tlp}

\usepackage{latexsym}
\usepackage{harvard}

\newtheorem{theorem}{Theorem}
\newtheorem{lemma}[theorem]{Lemma}
\newtheorem{proposition}[theorem]{Proposition}
\newtheorem{example}[theorem]{Example}
\newtheorem{definition}[theorem]{Definition}
\newtheorem{corollary}[theorem]{Corollary}
\newtheorem{remark}[theorem]{Remark}

\newcommand{\la}{\mbox{$\:\leftarrow\:$}}

\newenvironment{program}{\tt \begin{tabbing}pro\= {\tt pro}\= clause \kill}{\end{tabbing}}

\newenvironment{program2}{\tt \begin{tabbing}pro\= {\tt c1:} \= head\= hereisthebodyoftheclause  \= \kill}{\end{tabbing}}

\newcommand{\ol}[1]{{\bf #1}}

\title[Theory and Practice of Logic Programming]
      {On Modular Termination Proofs of \\
General Logic Programs}

\author[A. Bossi, N. Cocco, S. Etalle, and S. Rossi]
{ANNALISA BOSSI, NICOLETTA COCCO, SABINA ROSSI\\
Dipartimento di Informatica,
  Universit\`a Ca' Foscari di  Venezia \\
  via Torino 155,
  30172 Venezia, Italy
\and
SANDRO ETALLE\\
Department of Computer Science, University of Twente\\
P.O. Box 217, 7500 AE Enschede, The Netherlands\\
and\\
  CWI -- Center for Mathematics and Computer Science,\\
  P.O.\ Box 94079, 1090 GB Amsterdam, The Netherlands}

\begin{document}
\maketitle

\begin{abstract}      
\noindent      
We propose a modular method for proving termination of   
general logic programs (i.e., logic programs with negation). 
It is
 based on the notion of acceptable programs, but
it allows us to prove termination in a truly modular way.    
We consider programs consisting of a hierarchy of modules  
and supply a general result for proving  
termination by dealing with each module separately.  
For programs which are in a certain sense
well-behaved,
namely well-moded or well-typed programs,  
we derive both a simple verification technique and   
an iterative proof method.  
Some examples show how our system allows  
for greatly simplified proofs.  
\end{abstract}

\section{Introduction}      
    
It is standard practice to tackle a large proof by decomposing it into  
more managea\-ble pieces (lemmata or modules) and proving them  
separately.  By composing appropriately these simpler results, one can  
then obtain the final proof.  This methodo\-logy has been recognized an  
important one also when proving termination of logic programs.
Moreover most practical logic programs are  
engineered by assembling different modules and libraries, some of  
which might be pre-compiled or written in a different programming  
language. In such a situation, a compositional methodology for proving  
termination is of crucial importance. 

The first approach to modular termination proofs of logic programs
has been proposed
by Apt and Pedreschi in \cite{AP94}. It extends the seminal work
on \emph{acceptable} programs  \cite{AP93} which provides
 an algebraic characterization   
of programs terminating under Prolog left-to-right selection rule.
The class of acceptable programs 
contains  programs which terminate on  ground queries.
To prove acceptability one needs to determine a measure
on literals (\emph{level mapping}) such that,
 in any clause, the measure of the head is greater than the measure
of  each body literal. 
This implies the decreasing of  the measure of the literals resolved
during any computation starting from a ground or \emph{bounded}
query and hence  termination.

The significance of 
a modular approach to termination of logic programs has been
recognized also by other authors; more recent proposals can be found in
\cite{PR96,Mar96,VSD99,EBC99,VSD01}.

All previous proposals (with the exception of \cite{VSD99,EBC99})
require the existence of a relation between the level mappings used 
to prove acceptability
of distinct modules. This is not completely satisfactory:
it would be nice to be able to put together modules which
were independently proved terminating, and be sure that
the resulting program is still terminating.

We propose a modular approach to termination which allows one
to reason  independently on each single module and get a termination 
result on the whole program.
We consider general logic programs, i.e., logic programs with negation,
 employing SLDNF-resolution  
together with the leftmost selection rule
(also called \emph{LDNF-resolution}) as computational mechanism.  
We consider programs which can be divided into modules in a  
hierarchical way, so that each module is an extension of the previous  
ones.  We show that in this context the termination proof of the entire  
program can be given in terms of separate proofs for each module,  
which are naturally much simpler than a proof for the whole program.  
While assuming a hierarchy still allows one  
to tackle most real-life programs, it leads to termination proofs  
which, in most cases, are extremely simple.

 We characterize the class of queries terminating for the
whole program by introducing a new  notion of  
boundedness, namely \emph{strong boundedness}.
Intuitively, strong boundedness
 captures the queries which preserve
(standard) boundedness through the computation.
   By proving acceptability of each module wrt.\
a level mapping which measures only the predicates
defined in that module,
 we get a termination result for  
the whole program which is valid for any strongly bounded query. 
Whenever the original program is decomposed into a 
 hierarchy of small modules, the termination proof
can be drastically simplified with respect to
previous modular approaches.
Moreover strong boundedness can be naturally guaranteed by  
common persistent properties of programs and queries, namely properties  
preserved  through  LDNF-resolution such as 
\emph{well-modedness} \cite{DM85}
or \emph{well-typedness} \cite{BLR92}.

The paper is organized as follows.  Section \ref{prel} contains some 
preliminaries.  In particular we briefly recall the key concepts of  
LDNF-resolution, accepta\-bility, boundedness and program extension.  
Section \ref{R-acc} contains our main results which show how  
termination proofs of separate programs can be combined to obtain  
proofs of larger programs.  In particular we define the concept of  
strongly bounded query and we prove that for general programs composed  
by a hierarchy of $n$ modules, each one independently acceptable wrt.\   
its own level mapping, any strongly bounded query terminates.  In  
Section \ref{sec:applications} we show how strong boundedness is naturally ensured  
by some program properties which are preserved
 through   LDNF-resolution such as  
well-modedness and well-typedness.  In Section \ref{sec:iteration} we  
show how these properties allow us to apply our general results also for  
proving termination of modular programs in an iterative way.  In Section  
\ref{sec:comparisons} we compare our work with Apt and Pedreschi's approach.
Other related works and concluding remarks are discussed in 
Section \ref{conclusion}.
      
\section{Preliminaries}      
\label{prel}      
      
We use standard notation and terminology of logic programming
\cite{Llo87,Apt90,Apt97}. Just note that general logic programs are called      
in \cite{Llo87} normal logic programs.        
      
\subsection{General Programs and LDNF-Resolution}      
A \emph{general clause} is a construct of the form  
$$H\leftarrow L_1,\ldots, L_n$$  
with $(n\geq 0)$, where $H$ is an atom and $L_1,\ldots,L_n$ are 
literals (i.e.,  
either  atoms or the negation of  atoms).  
In turn, a \emph{general query} is a      
possibly empty finite sequence of literals $ L_1,\ldots,L_n$,  
with ($n\geq   0$).  
     A \emph{general program} is a finite set of  
general clauses\footnote{In the examples through the paper, we will  
  adopt the syntactic conventions of Prolog so that each query and  
  clause ends with the period ``.'' and ``$\leftarrow$'' is omitted in  
  the unit clauses.}.  
Given a query $Q:=L_1,\ldots, L_n$, a \emph{non-empty prefix of}
 $Q$ is any 
query $ L_1,\ldots, L_i$ with $i\in\{1,\ldots,n\}$.
For a literal $L$, we denote by ${\it   rel}(L)$ the predicate symbol of $L$.

Following the convention      
adopted  in \cite{Apt97}, we use bold characters to denote      
sequences of objects (so that \textbf{L} indicates a sequence of      
literals $L_1,\ldots,L_n$, while \textbf{t} indicates a sequence of      
terms $t_1,\ldots,t_n$).      
  
 For a given program $P$, we use the following  
notations: $B_P$ for the Herbrand base of $P$, ${\it ground}(P)$ for  
the set of all ground instances of clauses from $P$, ${\it comp}(P)$  
for the Clark's completion of $P$ \cite{Cla78}.

 Since in this paper we deal with general queries, clauses and programs,  
 we omit from now on the qualification ``general'', unless some  
 confusion might arise.

 We consider \emph{LDNF-resolution},  
and following Apt and Pedreschi's approach  
in studying the termination of  
general  programs \cite{AP93}, we view  LDNF-resolution as 
a top-down  
interpreter which, given a general program $P$ and a general  
query  $Q$, attempts to build a search tree for $P\cup\{Q\}$  
by constructing its branches in parallel.  
The branches in this tree are called \emph{LDNF-derivations}  
of $P\cup\{Q\}$ and the tree itself is called  \emph{LDNF-tree}  
of $P\cup\{Q\}$.  
Negative literals are resolved using the {\em negation-as-failure} rule  
which calls for the construction of a \emph{subsidiary LDNF-tree}.  
If during this subsidiary construction the interpreter diverges, the (main)  
LDNF-derivation 
is considered to be infinite.      
An LDNF-derivation is finite also if during its construction  
the interpreter encounters a query with the first literal  
being negative and non-ground.   
In such a case we say that the LDNF-derivation {\em flounders}.  
  
By termination of a general  program we actually mean  
termination of the underlying interpreter.  
Hence in order to ensure termination of a query $Q$  
in a program $P$, we require that all  LDNF-derivations of $P \cup   
\{Q\}$ are finite.

By an  \emph{LDNF-descendant} of  $P\cup\{Q\}$  we mean  
any query occurring during the  LDNF-resolution of  $P\cup\{Q\}$,  
including $Q$   
and all the queries occurring during the  
construction of the subsidiary LDNF-trees for $P\cup\{Q\}$.

For a non-empty query $Q$,  
we denote by $\mathit{first}(Q)$ the first literal of $Q$.    
Moreover we define $\mathit{Call}_P(Q)=\{ \mathit{first}(Q')\ |\  
Q' \mbox{ is an LDNF-descendant of }   P\cup\{Q\}\}$.  
It is worth noting that if $\neg A\in\mathit{Call}_P(Q)$ and $A$  
is a ground atom, then $A \in\mathit{Call}_P(Q)$ too.  
Notice that, for definite programs, the set $\mathit{Call}_P(Q)$ coincides
with the call set $\mathit{Call}(P,\{Q\})$ in \cite{DVB92,DDV99}.

The following trivial proposition holds.  
  
\begin{proposition}  
\label{prop-call}  
Let $P$ be a program and $Q$ be a query.  
All  LDNF-derivations of $P\cup\{Q\}$ are finite   
iff  for all positive literals  
$A\in \mathit{Call}_P(Q)$,  
all  LDNF-derivations of $P\cup\{A\}$ are finite.  
\end{proposition}

\subsection{Acceptability and Boundedness}  
     
The method we are going to use for proving  termination of modular programs  
is based on the concept of {\em acceptable}
 program \cite{AP93}. In order to introduce it, we  
start by the following definition, originally due to  \cite{Bez93} and  
 \cite{Cav89}.  
      
\begin{definition}[Level Mapping]      
  A \emph{level mapping} for a program $P$ is a function      
  $|\ |:B_P\rightarrow {\bf N}$ of ground atoms to natural numbers.      
By convention, this definition is extended in a natural way to      
  ground literals by putting $|\neg A|=|A|$.      
 For a ground literal $L$, $|L|$ is called the \emph{level} of      
  $L$.        
\end{definition}      

We will use the following notations.  
  Let $P$ be a program and $p$ and $q$ be relations.  
We say that $p$ \emph{refers to} $q$ if there is a clause in      
  $P$ that uses $p$ in its head and $q$ in its body;      
 $p$ \emph{depends on} $q$ if $(p,q)$ is in the      
  reflexive, transitive closure of the relation \emph{refers to}.    
We say that $p$ and $q$ are \emph{mutually recursive} and write   
$p\simeq q$, if $p$ depends on $q$ and $q$ depends on $p$.  
We also write $p\sqsupset q$, when $p$ depends on $q$ but $q$  
does not depend on $p$.  
  
We denote by ${\it Neg}_P$  the set of relations in $P$ which occur in      
  a negative literal in a  clause of $P$ and by   
 ${\it Neg}^*_P$  the set of relations in $P$ on which the      
  relations in ${\it Neg}_P$ depend.      
$P^-$ denotes the set of clauses in $P$ defining a relation of  
  ${\it Neg}^*_P$.

In the sequel we refer to the standard definition of model of a      
program and model of the completion of a program, see      
\cite{Apt90,Apt97} for details. In particular we need the      
following notion of {\em complete model} for a program.      
\begin{definition}[Complete Model]      
  A model $M$ of a program $P$ is called \emph{complete} if its      
  restriction to the relations from ${\it Neg}^*_P$ is a model of      
  ${\it comp}(P^-)$.        
\end{definition}      
Notice that  if $I$ is a model of ${\it comp}(P)$ then
its restriction to the relations in ${\it Neg}^*_P$ is a model
of ${\it comp}(P^-)$; hence $I$ is a complete model
 of $P$.

The following notion of acceptable program was introduced in \cite{AP93}.     
Apt and Pedreschi  proved that such a notion  
fully characterizes left-termination, namely termination wrt.\ any ground   
query, both for definite programs and for general programs which   
have no LDNF-derivations which flounder.

\begin{definition}[Acceptable Program]      
  Let $P$ be a program, $|\ |$ be a level mapping for $P$ and $M$ be a      
  complete model of $P$.      
$P$ is called \emph{acceptable wrt.\  $|\ |$ and $M$}  if for every      
 clause $A\leftarrow {\bf A}, B,{\bf B}$ in ${\it ground}(P)$    
the following implication holds:  
  $$ \mbox{ if } M\models {\bf A} \mbox{ then } |A|>|B|. $$  
  
\end{definition}  
Note that  if $P$ is a definite program, then both $P^-$
and ${\it Neg}^*_P$ are empty and $M$ can be 
any model of $P$.

We also need the notion of bounded atom.  
  
\begin{definition}[Bounded Atom]      
\label{def-Boundedness}      
Let $P$ be a program and $|\ |$ be a level mapping for $P$.  
An atom $A$ is called \emph{bounded wrt.\    $|\ |$}  
if the set of all $|A'|$, where $A'$ is a ground instance of $A$, is finite.  
In this case we   
denote by ${\it max}|A|$ the maximum value in this set.  
\end{definition}  
  
Notice that if an atom $A$ is bounded then, by definition of level   
mapping, also the corresponding negative literal, $\neg A$,  
is bounded.   
  
Note also that, for atomic queries,
 this definition coincides with the definition  
of bounded query introduced in \cite{AP93}  
in order to characterize terminating queries for acceptable
programs. 
In fact, in  case of atomic queries the notion of boundedness does  
not depend on a model.

\subsection{Extension of a Program}      

In this paper we consider a hierarchical situation where a program  
uses another one as a subprogram. The following definition 
formalizes this situation.  
      
\begin{definition}[Extension]      
Let $P$ and $R$ be two  programs.      
A relation $p$ is \emph{defined in} $P$ if $p$      
  occurs in a head of a clause of $P$;     
  a literal $L$ is      
  \emph{defined in} $P$ if ${\it rel}(L)$ is defined in $P$;      
 $P$ \emph{extends} $R$, denoted      
 $P\sqsupset R$, if no relation defined in $P$      
  occurs in~$R$.             
\end{definition}      
      
Informally, $P$ extends $R$ if $P$ defines new relations with respect to      
 $R$. 
Note that $P$ and $R$ are independent if no relation defined in $P$
occurs in $R$ and  no relation defined in $R$
occurs in $P$, i.e. $P\sqsupset R$ and $R\sqsupset P$.

In  the sequel we will study termination  in a hierarchy of 
programs.

\begin{definition}[Hierarchy of Programs] 
Let $P_1,\ldots, P_n$ be programs such that  for all $i\in\{1,\dots,n-1\}$,
$P_{i+1} \sqsupset (P_1\cup
\cdots \cup P_{i})$.
Then we call  $P_n \sqsupset \cdots  \sqsupset P_1$ 
 a \emph{hierarchy of
programs}.
\end{definition}

\section{Hierarchical Termination}      
\label{R-acc}      

This section contains our main results which show how termination  
proofs of separate programs can be combined to obtain proofs of larger  
programs.    
We start with a technical result, dealing with the case in  
which a program consists of a hierarchical combination of two  
modules. This is the base both of a generalization to a hierarchy  
of $n$ programs and of an iterative proof method for termination   
presented in Section~\ref{sec:iteration}.  
Let us first introduce the following notion of {\em $P$-closed}
 class of queries. 
  
 \begin{definition}  [P-closed Class]    
  Let ${\cal C}$ be a class of queries and $P$ be a program.  
 We say that ${\cal C}$ is {\em  
    $P$-closed} if it is   closed under non-empty prefix (i.e.,   
 it contains all the non-empty prefixes of its elements) and for each query  
$Q\in {\cal C}$,  
every LDNF-descendant of $P\cup\{Q\}$ is contained in  ${\cal C}$.  
\end{definition}    
  
Note that if ${\cal C}$  is $P$-closed,   
then for each query  $Q \in {\cal C}$,   
$\mathit{Call}_{P}(Q) \subseteq {\cal C}$.  
  
We can now state our first general theorem.  
  Notice that if $P$  extends $R$ and $P$ is  acceptable      
wrt.\  some level mapping $|\ |$ and model $M$,  then  
$P$ is  acceptable   also   
wrt.\  the level mapping $|\ |'$ and $M$, where
    $|\ |'$ is defined on      
the  Herbrand base of the union of the two programs $B_{P\cup R}$  
and  it  takes the value      
$0$ on  the literals which are not defined in $P$      
(and hence, in particular,  on  the literals which occur in $P$ but      
are  defined in $R$). This shows that in each module
it is sufficient to compare  only the level of the literals defined inside it,
while we can ignore literals defined outside the module. 
In the following we make use of this observation in order
to associate to each module in a hierarchy a level mapping
which is independent from the context.
    
\begin{theorem}      
    \label{theo:generale-sulle-classi}      
    Let $P$ and $R$ be two programs such that $P$ extends $R$, $M$ be  
    a complete model of $P\cup R$ and ${\cal C}$ be a  $(P\cup R)$-closed
 class of queries.
  Suppose that  
\begin{itemize}      
\item $P$ is acceptable wrt.\  a  level mapping $|\ |$ and $M$,      
\item for all queries $Q\in {\cal C}$, all LDNF-derivations of $R\cup\{Q\}$  
are finite,  
\item for all  atoms $A\in {\cal C}$, if $A$ is defined in $P$  
  then $A$ is bounded wrt.\ $|\ |$.  
\end{itemize}  
Then  for all queries  $Q\in {\cal C}$,   
  all  LDNF-derivations of $(P\cup R)\cup\{Q\}$ are finite.  
\end{theorem}     
\begin{proof}  
By the fact that  ${\cal C}$ is   $(P\cup R)$-closed and Proposition  
\ref{prop-call}, it is sufficient to prove that for all positive literals  
$A\in  {\cal C}$,  
all  LDNF-derivations of $(P\cup R)\cup\{A\}$ are finite.  
Let us consider an  atom  $A\in {\cal C}$.  
  
If $A$ is defined in $R$, then the thesis trivially holds by   
hypothesis.  
  
If $A$ is defined in $P$,  
 $A$ is bounded wrt.\  $|\ |$  by hypothesis   
 and thus  $\mathit{max}|A|$ is defined.  
The proof proceeds   
 by induction on $\mathit{max}|A|$.  
  
\emph{Base}.   
Let $\mathit{max}|A|=0$.  
 In this case, by acceptability of $P$, there are no  
clauses in $P$ whose head unifies with $A$  
and whose body is non-empty.  
Hence, the thesis holds.

\emph{Induction step}.   
Let $\mathit{max}|A|>0$.   
It is sufficient to prove that for all direct descendants 
$(L_1,\ldots,L_n)$ in the LDNF-tree of $(P\cup R)\cup\{A\}$, 
if
$\theta_i$ is a computed answer for  $P\cup\{L_1,\ldots,L_{i-1}\}$  
then all  LDNF-derivations of   
$(P\cup R)\cup \{L_i\theta_i\}$ are finite.

Let $c: H'\leftarrow L'_{1},\ldots , L'_{n}$ be a clause of $P$ such that  
 $\sigma = mgu(H', A)$.  
Let  $H=H'\sigma$ and for all  
  $i\in \{1,\ldots,n\}$,  let $L_i=L'_i\sigma$  
and  $\theta_i$ be  a substitution such that  
$\theta_i$ is a computed answer of $L_1,\ldots,L_{i-1}$  
in $P\cup R$.

We distinguish two cases. If $L_i$ is defined in $R$ then the thesis  
follows by hypothesis.  
  
Suppose that $L_i$ is defined in $P$.  
We prove that $L_i\theta_i$ is bounded and $\mathit{max}|A|>  
\mathit{max}|L_i\theta_i|$. The thesis will follow by the  
induction hypothesis.

Let $\gamma$ be a substitution such that  $L_i\theta_i\gamma$ is ground.  
By soundness of LDNF-resolution \cite{Cla78},   
there exists $\gamma'$ such that   
 $M\models (L_1,\ldots, L_{i-1})\gamma'$ and
$c\sigma\gamma'$ is a ground instance of  
 $c$ and   
$L_i\gamma'=L_i\theta_i\gamma$. Therefore 
\[\begin{array}{lllll}    
|L_i\theta_i\gamma  | & = & |L_i\gamma'| &\\  
& = &  |L'_i\sigma \gamma'| & (\mbox{since } L_i=L'_i\sigma)\\ 
& < & |H'\sigma\gamma'|   & (\mbox{since $P$ is acceptable})\\ 
& = & |A\sigma\gamma'|  & (\mbox{since } \sigma = mgu (H', A)).  
\end{array}\]  
Since $A$ is bounded, we can conclude that $L_i\theta_i$ is bounded  
and also that  $\mathit{max}|A|> \mathit{max}|L_i  \theta_i|$.  
\end{proof}  
\medskip
  
We are going to extend the above theorem in order to handle the presence  
of more than two modules. We need to introduce more notation.  
Let us consider the case of a program $P$ consisting of a  
hierarchy  
$R_n\sqsupset \ldots \sqsupset R_1$ of distinct modules, and satisfying  
the property that each module, $R_i$, is acceptable wrt.\ a distinct  
level mapping, $|\ |_i$, and a complete model, $M$, of the whole  
program. Under these  
assumptions we identify a specific class of queries which terminate in  
the whole program.  We characterize the class of terminating queries  
in terms of the following notion of strong boundedness.  This class  
enjoys the property of being $P$-closed.  
       
\begin{definition}[Strongly Bounded Query]      
\label{def:strongn}     
Let the program $P:=R_1\cup \ldots \cup R_n$ be a hierarchy
$ R_n\sqsupset \ldots \sqsupset R_1$ and $|\ |_1,\ldots,|\ |_n$   
be level mappings for $R_1,\ldots,R_n$, respectively.  A query  
$Q$ is called \emph{strongly bounded wrt.\ $P$ and   
$|\ |_1,\ldots,|\ |_n$} if  
\begin{itemize}      
\item      
for all  atoms $A\in \mathit{Call}_P(Q)$,   
 if  $A$ is defined in $R_i$      
 (with $i\in \{1,\ldots,n\}$) then $A$  is bounded wrt.\ $|\ |_i$.       
\end{itemize}         
\end{definition}      
      
Notice that the notion of boundedness for an atom (see Definition  
\ref{def-Boundedness}) does not depend on the choice of a particular  
model of $P$.  As a consequence, also the definition of strong  
boundedness does not refer to any model of $P$; however, it refers to  
the LDNF-derivations of $P$. For this reason, a ground atom is always  
bounded but not necessarily strongly bounded.  On the other hand, if $A$  
is strongly bounded then it is bounded too.    
  
The following remark follows immediately.  
  
\begin{remark}      
\label{i-strong}      
Let the query $Q$ be strongly bounded wrt.\ $P$
and $|\ |_1,\ldots,|\ |_n$, 
where $P$ is a hierarchy $R_n \sqsupset \cdots  \sqsupset R_1$ .  
Let $i\in\{1,\ldots,n\}$.   
If $Q$ is defined in $R_1\cup \ldots \cup R_i$ then $Q$ is strongly bounded wrt.\   
$R_1\cup \ldots\cup R_i$ and $|\ |_{1},\ldots,|\ |_{i}$.  
\end{remark}      
  
In order to verify whether a query Q is strongly bounded wrt.\ a given  
program $P$ one can perform a call-pattern analysis  
\cite{JB92,GG94,CD95} which allows us to infer information about the  
form of the call-patterns, i.e., the atoms that will be possibly  
called during the execution of $P\cup \{Q\}$.    
However this is not the only way  
for guaranteeing strong boundedness.  There are classes of programs  
and queries for which strong boundedness can be proved in a  
straightforward way.  This is shown in the following section.    
  
Let us illustrate the notion of strong boundedness through an example.  
   
\begin{example}      
\label{exa:sb}       
Let \texttt{LIST01}
be the following program which defines 
the  proper lists  of  \texttt{0}'s and \texttt{1}'s, i.e.
lists 
 containing only  \texttt{0}'s and \texttt{1}'s and at least
two distinct elements, as follows:   

\begin{program2}      
\> r1: \> list01([ ],0,0).\\      
\> r2: \> list01([0|Xs],s(N0),N1) \la list01(Xs,N0,N1).\\      
\> r3: \> list01([1|Xs],N0,s(N1)) \la list01(Xs,N0,N1).\\[2mm]  
\> r4: \> length([ ],0).\\      
\> r5: \> length([X|Xs],s(N)) \la length(Xs,N).  \\[2mm] 
\> r6: \> plist01(Ls) \la  list01(Ls,N0,N1), \\
\> \> \> $\neg$length(Ls,N0), 
$\neg$length(Ls,N1). 
\end{program2} 
Let us  distinguish two modules in \texttt{LIST01}:
$R_1=\{\mathtt{r_1},\mathtt{r_2},\mathtt{r_3},\mathtt{r_4},\mathtt{r_5}\}$ and
$R_2=\{\mathtt{r_6}\}$ ($R_2$ extends $R_1$).
Let $|\ |_1$ be  the natural level mapping  for
$R_1$ defined by:      
      
\begin{program}      
\> \(\mathtt{|list01(\mathit{ls},\mathit{n0},\mathit{n1})|_1 = |\mathit{ls}|_{length}}\)\\  
  \> \(\mathtt{|length(\mathit{ls},\mathit{n})|_1 = |\mathit{n}|_{size}}\)  
\end{program}      
where for a term $\mathit{t}$, if $\mathit{t}$ is a list then      
 \(\mathtt{|\mathit{t}|_{length}}\)  is equal to      
the length of the list, otherwise it is $0$, while     
 \(\mathtt{|\mathit{t}|_{size}}\)      
 is the number of function symbols occurring in the term  $\mathit{t}$.
%
%
Let  also $|\ |_2$ be  the trivial level mapping  for $R_2$ defined by:      
      
\begin{program}      
\> \(\mathtt{|plist01(\mathit{ls})|_2 = 1}\) 
\end{program}      
and assume that  $|L|_2 = 0$, if  $L$ is not defined in $R_{2}$.  
    
Let us consider the following sets of atomic queries for   
$\mathtt{LIST01}:=R_1\cup R_2$:  
     
\begin{program}
 \({\cal Q}_1 \ =\ \{\mathtt{list01(\mathit{ls},\mathit{n0},\mathit{n1})} | \      
\mathit{ls} \textrm{ is a list,  possibly non-ground,  of a fixed length}\};\)\\[1mm]      
 \({\cal Q}_2\ =\ \{\mathtt{length(\mathit{ls},\mathit{n})}  | \ \mathit{n}      
\textrm{  is a ground term of the form  either 0  or } \texttt{s(s(\ldots(0)))} \};\)\\[1mm]  
 \({\cal Q}_3 \ =\ \{\mathtt{plist01(\mathit{ls})} | \      
\mathit{ls} \textrm{ is a list,  possibly non-ground,  of a fixed length}\}\). 
\end{program}      

By definition of $|\ |_1$, all the atoms in ${\cal Q}_1$ and
 ${\cal Q}_2$ are    
bounded wrt.\  $|\ |_1$. Analo\-gously, all the atoms in  ${\cal Q}_3$   
are  bounded wrt.\  $|\ |_2$.    
Notice that  for all  atoms 
 $A\in {\it Call}_P({\cal Q}_j)$, 
with $j\in\{1,2,3\}$, there exists $k\in\{1,2,3\}$ such that
 $A\in {\cal Q}_k$.   Hence,
 if $A$ is defined in $R_i$ then $A$ is  
bounded wrt.\  $|\ |_i$.  
This proves that  the set of queries    
${\cal Q}_1$,  ${\cal Q}_2$    and ${\cal Q}_3$    
are strongly bounded wrt.\    
$\mathtt{LIST01}$ and  $|\  |_1$, $|\ |_2$.    
\end{example}

Here we introduce our main result.  
      
\begin{theorem}      
\label{theo:strongn}      
Let   $P:=R_1\cup \ldots \cup R_n$ be a program such that      
 $ R_n\sqsupset \ldots \sqsupset R_1$ is a hierarchy,       
$|\ |_1,\ldots,|\ |_n$   be level mappings for $R_1,\ldots,R_n$,      
respectively,       
and $M$ be a complete model of $P$.      
Suppose that      
\begin{itemize}      
\item       
 $R_i$ is acceptable wrt.\  $|\ |_i$ and $M$,   
for all $i\in\{1,\ldots,n\}$.      
\item $Q$  is a query  strongly bounded  wrt.\       
 $P$ and $|\ |_1,\ldots,|\ |_n$.      
\end{itemize}      
Then all  LDNF-derivations of $P\cup \{Q\}$      
are finite.      
\end{theorem}      
\begin{proof}            
Let $Q$ be a query  strongly bounded wrt.\  $P$ and      
  $|\ |_{1},\ldots,|\ |_n$.  
We prove the theorem by induction on $n$.  
   
\textit{Base}.   
Let $n=1$.  
This case follows immediately by Theorem \ref{theo:generale-sulle-classi},  
where $P=R_1$, $R$ is empty and ${\cal C}$ is the class of strongly  
bounded queries wrt.\  $R_1$ and $|\ |_1$, and the fact that  
a strongly bounded atom is also bounded.  
      
\emph{Induction step}.   
Let $n>1$.      
Also this case follows  by Theorem \ref{theo:generale-sulle-classi},  
where $P=R_n$, $R=R_1\cup \ldots \cup R_{n-1}$ and  
 ${\cal C}$ is the class of strongly  
bounded queries  wrt.\       
$R_1\cup \ldots \cup R_n$ and $|\ |_1,\ldots,|\ |_n$. In fact,  
\begin{itemize}  
\item  
 $R_n$ is acceptable wrt.\  $|\ |_n$ and $M$;  
\item   
for all queries $Q\in {\cal C}$, all LDNF-derivations of $(R_1\cup \ldots \cup R_{n-1})\cup\{Q\}$  
are finite, by Remark \ref{i-strong} and the inductive hypothesis;  
\item    
for all  atoms $A\in {\cal C}$,  
if $A$ is defined in $R_n$ then $A$ is bounded wrt.\  $|\ |_n$,  
by definition of strong boundedness.  
\end{itemize}  
\end{proof}  

Here are a few examples applying Theorem \ref{theo:strongn}.  
\begin{example}    
\label{exa:teo1}   
Let us reconsider the program of Example \ref{exa:sb}.    
In the program {\tt LIST01}, $R_1$ and $R_2$ are    
acceptable wrt.\  any complete model and     
the level mappings $|\ |_1$ and $|\ |_2$, respectively.   
We already showed that  ${\cal Q}_1, {\cal Q}_2$ and  
${\cal Q}_3$ are strongly bounded wrt.\  ${\tt LIST01}$ and
 $|\ |_1$, $|\ |_2$.
Hence, by  Theorem  \ref{theo:strongn},   
all LDNF-derivations of ${\tt LIST01}\cup\{Q\}$,  
 where $Q$ is a query in ${\cal Q}_1, {\cal Q}_2$ or   
${\cal Q}_3$, are finite.  
\end{example}
Notice that in the previous example the top module in the hierarchy,
$R_2$, contains no recursion. Hence it is intuitively clear that any 
problem for termination cannot depend on it. This is reflected by the 
fact that the level mapping for $R_2$ is completely trivial.
This shows how the hierarchical decomposition of the program can
 simplify the termination proof.
\begin{example}    
\label{exa:teo2}     
Consider the sorting program \texttt{MERGESORT} \cite{Apt97}:      
\begin{program2}      
  \> c1: \> mergesort([ ],[ ]).\\      
  \> c2: \> mergesort([X],[X]). \\      
  \> c3: \> mergesort([X,Y|Xs],Ys) \la \\      
 \> \> \> split([X,Y|Xs],X1s,X2s),\\      
  \> \> \> mergesort(X1s,Y1s),\\      
  \> \> \> mergesort(X2s,Y2s),\\      
  \> \> \> merge(Y1s,Y2s,Ys).\\[2mm]      
  \> c4: \> split([ ],[ ],[ ]).\\      
  \> c5: \> split([X|Xs],[X|Ys],Zs) \la   split(Xs,Zs,Ys).\\[2mm]      
  \> c6: \> merge([ ],Xs,Xs).\\      
  \> c7: \> merge(Xs,[ ],Xs).\\      
  \> c8: \> merge([X|Xs],[Y|Ys],[X|Zs]) \la      
  X<=Y, merge(Xs,[Y|Ys],Zs).\\      
  \> c9: \> merge([X|Xs],[Y|Ys],[Y|Zs]) \la X>Y, merge([X|Xs],Ys,Zs).      
\end{program2}      
      
Let us divide the program \texttt{MERGESORT} into three      
modules, $R_1,R_2,R_3$, such that $ R_3 \sqsupset R_2 \sqsupset R_1$      
as follows:      
      
\begin{itemize}      
\item $R_3:=\{\mathtt{c1}, \mathtt{c2}, \mathtt{c3}\}$, it defines the relation      
\texttt{mergesort},      
\item $R_2:=\{\mathtt{c4}, \mathtt{c5}\}$, it      
defines the relation      
\texttt{split},      
\item $R_1:=\{\mathtt{c6}, \mathtt{c7}, \mathtt{c8}, \mathtt{c9}\}$, it        
defines the relation      
\texttt{merge}.      
\end{itemize}          
    
Let us consider the natural level mappings    
        
\begin{program}      
\> \(\mathtt{|merge(\mathit{xs},\mathit{ys},\mathit{zs})|_{1} = |\mathit{xs}|_{length}+|\mathit{ys}|_{length}}\)\\[2mm]      
\> \(\mathtt{|split(\mathit{xs},\mathit{ys},\mathit{zs})|_{2} = |\mathit{xs}|_{length}}\)  \\[2mm]    
\> \(\mathtt{|mergesort(\mathit{xs},\mathit{ys})|_{3} = |\mathit{xs}|_{length}}\)     
\end{program}      
and assume that for all $i\in\{1,2,3\}$,      
\(\mathtt{|\textit{L}|_{i} = 0 \textrm{ if \textit{L} is not     
defined in }} R_i\).    
      
All  ground queries      
 are strongly bounded wrt.\  the program \texttt{MERGESORT} and      
the level mappings  $|\ |_1,|\ |_2,|\ |_3$.      
  Moreover, since the program is a definite one,
 $R_1$ and $R_2$ are    
acceptable wrt.\  any model and     
the level mappings $|\ |_1$ and $|\ |_2$,
respectively, while
$R_3$ is acceptable wrt.\  the level mapping  $|\ |_3$          
and the  model $M$ below:      
      
\begin{program2}      
 \> \(\mathit{M =}\)     \>  
\(\mathtt{[mergesort(Xs,Ys)]} \cup \mathtt{[merge(Xs,Ys,Zs)]}  \cup \)\\      
  \> \> \(\mathtt{\{split([\;],[\;],[\;])\}}  \cup \)\\      
  \> \> \(\mathtt{\{split([\mathit{x}],[\;],[\mathit{x}])|\   
\mathit{x} \textrm{ is any ground term}\}} \cup  \)\\      
   \> \> \(\mathtt{\{split([\mathit{x}],[\mathit{x}],[\;])|\      
\mathit{x} \textrm{ is any ground term}\}} \cup \)\\      
   \>\> \(\mathtt{\{split(\mathit{xs},\mathit{ys},\mathit{zs})|\ 
\mathit{xs}, \mathit{ys}, \mathit{zs} \textrm{ are ground terms and}   } \)\\  
 \> \> \(
 |\mathit{xs}|_{length}\geq 2,\;      
|\mathit{xs}|_{length}>|\mathit{ys}|_{length},      
  \mathtt{|\mathit{xs}|_{length}>      
|\mathit{zs}|_{length}\}}\)      
\end{program2}     
where    we denote by $[A]$ the set of all ground    
instances of an atom $A$.   
    
Hence, by  Theorem  \ref{theo:strongn},  
all LDNF-derivations of  
 ${\tt MERGESORT}\cup\{Q\}$, where $Q$ is a ground query, are finite.  
     
Note that by exchanging the roles of $R_{1}$ and $R_{2}$  
we would obtain the same result. In fact the definition of   
${\tt merge}$ and ${\tt split}$ are independent from each other.    
\end{example}

\section{Well-Behaving Programs}      
\label{sec:applications}      
      
In this section we consider the problem of how to prove that a     
query is strongly bounded. In fact one could argue that checking strong     
boundedness is more difficult and less abstract than checking boundedness     
itself in the sense of \cite{AP93}:    
we have to refer to all LDNF-derivations instead of referring to a model, which    
might well look like a step backwards in the proof of termination of a    
program. This is only partly true: in order to check strong  boundedness    
we can either  employ tools based on abstract interpretation     
or concentrate our attention only on programs   
which exhibit  useful
persistence properties wrt.\ LDNF-resolution.    
  
We now show how the well-established notions of well-moded and   
well-typed programs can be employed in order to verify strong boundedness  
and how they can lead to simple termination proofs.  
  
\subsection{Well-Moded Programs}  
    
The concept of a well-moded program is  due to  \cite{DM85}. The formulation we use here is from    
 \cite{Ros91}, and it is equivalent to that in   
\cite{Dra87}.  The original definition  was given for definite programs    
(i.e., programs without negation), however it applies to general    
programs as well, just by considering literals instead of atoms.    
It relies on the concept of  \emph{mode}, which is a function that     
labels the positions     
of each predicate in order to indicate how the arguments of a     
predicate should be used.    
    
\begin{definition}[Mode]    
  Consider an $n$-ary predicate symbol $p$.  By a \emph{mode} for $p$    
  we mean a function $m_p$ from $\{1,\ldots,n\}$ to the set $\{+,-\}$.    
  If $m_p(i)=+$ then we call $i$ an {\it input position} of $p$;    
  if $m_p(i)=-$ then we call $i$ an {\it output position} of $p$.    
  By a  \emph{moding} we mean  a collection of modes, one for each   
   predicate symbol.    
\end{definition}    
    
In a moded program, we assume that each predicate symbol has a unique     
mode associated to it.    
Multiple moding may be obtained by simply  renaming the predi\-cates.    
We use the notation \(p(m_p(1),\ldots,m_p(n))\) to denote the moding    
associated with a predicate $p$    
(e.g., $\mathtt{append(\textrm{+},\textrm{+},\mathrm{-})}$).    
Without loss of generality, we assume, when writing a literal  as   
$p({\bf s},{\bf t})$, that we are    
indicating with ${\bf s}$ the sequence of terms filling in the input    
positions of $p$ and with ${\bf t}$ the sequence of terms filling in    
the output positions of $p$.    
Moreover, we adopt the convention that $p({\bf s},{\bf t})$ 
could denote both negative and positive literals.

\begin{definition}[Well-Moded]    
\label{def:well-moded}    
\begin{itemize}    
\item    
A query $p_1({\bf s}_1,{\bf t}_1),\ldots,p_n({\bf s}_n,{\bf t}_n)$    
is called \emph{well-moded} if  for all $i\in\{1,\ldots,n\}$    
$${\it Var}({\bf s}_i)\subseteq \bigcup_{j=1}^{i-1}{\it Var}({\bf    
  t}_j).$$    
\item    
A clause $p({\bf t}_0,{\bf s}_{n+1})\leftarrow     
p_1({\bf s}_1,{\bf t}_1),\ldots,p_n({\bf s}_n,{\bf t}_n)$    
is called \emph{well-moded} if  for all $i\in\{1,\ldots,n+1\}$    
$${\it Var}({\bf s}_i)\subseteq \bigcup_{j=0}^{i-1}{\it Var}({\bf t}_j).$$    
\item A program is called \emph{well-moded} if all of its clauses are    
  well-moded.      
\end{itemize}    
\end{definition}    
    
    
Note that well-modedness can be syntactically checked  
in a time which is linear wrt.\ the size of the program (query).  
  
\begin{remark}  
\label{remark:prefix_wm}  
If $Q$ is a well-moded query then all its prefixes are well-moded.  
\end{remark}

The following lemma states that well-moded queries are closed under     
LDNF-resolution. This result has been proved in \cite{AP94a}    
for LD-derivations and definite programs.  
    
\begin{lemma}  
\label{lemma:well-moded-closed}    
Let $P$ and $Q$ be a well-moded program and query, respectively. Then all   
 LDNF-descendants of $P\cup\{Q\}$ are well-moded.  
\end{lemma}    
\begin{proof}  
It is sufficient to extend the proof in \cite{AP94a}  
by showing that if a query $\neg A, L_1,\ldots,L_n$   
is  well-moded  and $A$ is ground  
then both $A$ and $L_1,\ldots,L_n$ are well-moded.  
This follows immediately by definition of well-modedness.  
If $A$ is non-ground then the query above has no descendant.  
\end{proof}

When considering well-moded programs, it is natural to   
measure atoms only in their input positions \cite{EBC99}.  
  
\begin{definition}[Moded Level Mapping]  
  Let $P$ be a moded program. A function $|\;|$ is a \emph{moded  
    level mapping for $P$} if it is a level mapping for $P$  
such that  
\begin{itemize}  
\item for any $\mathbf{ s}$, $\mathbf{ t}$ and $\mathbf{ u}$, $|p(\mathbf{  
    s},\mathbf{ t})|=|p(\mathbf{ s},\mathbf{ u})|$.  
\end{itemize}  
\end{definition}  
  
Hence in a moded level mapping
 the level of an atom is independent from the terms  
in its output positions.

The following Remark and Proposition allow us to exploit well-modedness  
for applying Theorem \ref{theo:strongn}.

\begin{remark}  
\label{remark-modedlevelmapp}  
Let $P$ be a well-moded program.  
If $Q$ is well-moded, then $\emph{first}(Q)$ is ground in its input      
position and hence it is bounded wrt.\   any moded level mapping for $P$.  
Moreover,  by      
Lemma \ref{lemma:well-moded-closed},  
every well-moded query  is strongly bounded wrt.\  $P$  
and any moded level mapping for $P$.  
\end{remark}  
      
\begin{proposition}      
\label{pro:strongbound_wm}      
Let   $P:=R_1\cup \ldots \cup R_n$ be a \emph{well-moded} program  
 and   
    $ R_n\sqsupset \ldots \sqsupset R_1$ a hierarchy,  and       
$|\ |_1,\ldots,|\ |_n$   be       
\emph{moded} level mappings for $R_1,\ldots,R_n$,      
respectively.\\      
 Then every well-moded query is      
strongly bounded wrt.\  $P$ and $|\ |_1,\ldots,|\ |_n$.      
\end{proposition}      
  
\begin{example}    
\label{exa:well-moded-program1}    
Let \texttt{MOVE} be the following program which defines
a permutation between two lists such that only one element is moved.
We introduce  modes and we distinguish the two uses of \texttt{append}
by renaming it as  \texttt{append1} and \texttt{append2}.
\begin{program2}      
 \> mode  \(\mathtt{delete(\mathit{+},\mathit{-},\mathit{-})}\).\\    
 \> mode  \(\mathtt{append1(\mathit{-},\mathit{-},\mathit{+}}\)). \\      
 \> mode  \(\mathtt{append2(\mathit{+},\mathit{+},\mathit{-})}\).\\      
 \> mode  \(\mathtt{move(\mathit{+},\mathit{-})}\). 
\end{program2} 
\begin{program2}      
\>r1: \>  delete([X|Xs],X,Xs).\\      
\>r2: \>  delete([X|Xs],Y,[X|Ys]) \la delete(Xs,Y,Ys).  \\[2mm]    
\>r3: \>  append1([ ],Ys,Ys).\\      
\>r4: \>  append1([X|Xs],Ys,[X|Zs]) \la append1(Xs,Ys,Zs).  \\[2mm]  
\>r5: \>  append2([ ],Ys,Ys).\\      
\>r6: \>  append2([X|Xs],Ys,[X|Zs]) \la append2(Xs,Ys,Zs).  \\[2mm] 
\>r7: \>  move(Xs,Ys) \la  append1(X1s,X2s,Xs),\\      
\> \> \> delete(X1s,X,Y1s), append2(Y1s,[X|X2s],Ys). 
\end{program2}   
   Let us partition   \texttt{MOVE} into the modules
$R_1=\{\mathtt{r_1},\mathtt{r_2},\mathtt{r_3},\mathtt{r_4}, \mathtt{r_5},\mathtt{r_6}\}$ and
$R_2=\{\mathtt{r_7}\}$ ($R_2$ extends $R_1$).
Let $|\ |_1$ be
the natural level mapping  for $R_1$ defined by: 
     
\begin{program}      
  \> \(\mathtt{|append1(\mathit{xs},\mathit{ys},\mathit{zs})|_1 = |\mathit{zs}|_{length}}\)\\      
   \> \(\mathtt{|append2(\mathit{xs},\mathit{ys},\mathit{zs})|_1 = |\mathit{xs}|_{length}}\).  \\
    \> \(\mathtt{|delete(\mathit{xs},\mathit{x},\mathit{ys})|_1 = |\mathit{xs}|_{length}}\). 
\end{program}  
$R_2$ does not contain any recursive definition
hence
let $|\ |_2$ be the trivial level mapping defined by:      
  \begin{program}      
  \> \(\mathtt{|move(\mathit{xs},\mathit{ys})|_2 = 1}\)      
\end{program}      
and assume that      
$|L|_2 = 0$, if $L$ is not defined in $R_{2}$.

The program   
  ${\tt MOVE}:=R_1\cup R_2$ is well-moded    
and hence by Proposition \ref{pro:strongbound_wm}     
every  well-moded query   is strongly bounded wrt.\ {\tt MOVE} 
  and    
$|\ |_1$, $|\ |_2$.    
\end{example}

\begin{example}    
\label{exa:well-moded-program2}    
Let $R_1$ be the program       
which defines the relations \texttt{member} and    
\texttt{is},      
$R_2$ be the program  
defining the relation \texttt{count} and
$R_3$ be the program  
defining the relation
 \texttt{diff}  with the moding and the definitions below.      
\begin{program2}       
\> mode \(\mathtt{member(\mathit{+},\mathit{+})}\).\\      
\> mode \(\mathtt{is(\mathit{-},\mathit{+})}\).\\      
\> mode \(\mathtt{diff(\mathit{+},\mathit{+},\mathit{+},\mathit{-})}\).\\      
\> mode \(\mathtt{count(\mathit{+},\mathit{+},\mathit{-})}\).\\[2mm]

\> r1: \> member(X,[X|Xs]).\\      
\> r2: \> member(X,[Y|Xs]) \la member(X,Xs).  \\[2mm]   
\> r3: \>  diff(Ls,I1,I2,N) \la count(Ls,I1,N1),  count(Ls,I2,N2), \\      
\> \> \>  N is N1-N2.\\[2mm]      
\> r4: \> count([ ],I,0).\\      
\> r5: \> count([H|Ts],I,M) \la  member(H,I), count(Ts,I,M1), \\      
\> \> \>M is M1+1.\\      
\> r6: \> count([H|Ts],I,M) \la   \(\neg\) member(H,I), count(Ts,I,M).  
\end{program2}      
The relation      
$\mathtt{diff(\mathit{ls},\mathit{i1},\mathit{i2},\mathit{n})}$, given a list $\mathit{ls}$ and two check-lists      
$\mathit{i1}$ and $\mathit{i2}$, defines   
the difference $\mathit{n}$  between the number      
of elements of $\mathit{ls}$ occurring in $\mathit{i1}$ and the number of      
elements of $\mathit{ls}$ occurring in $\mathit{i2}$.  
Clearly $R_3\sqsupset R_2 \sqsupset R_1$.
It is easy to see      
that $R_1$ is     
acceptable wrt.\  any complete model and the moded level mapping    
    
\begin{program}      
  \> \(\mathtt{|member(\mathit{e},\mathit{ls})|_1 = |\mathit{ls}|_{length}}\)     
\end{program}       
    $R_2$ is acceptable wrt.\  any complete model  
and the  moded  level mapping:      
\begin{program}          
  \> \(\mathtt{|count(\mathit{ls},\mathit{i},\mathit{n})|_2 = |\mathit{ls}|_{length}}\)     
\end{program}      
and $R_3$ is acceptable wrt.\  any complete model  
and the  trivial moded  level mapping:  
\begin{program} 
  \> \(\mathtt{|diff(\mathit{ls},\mathit{i1},\mathit{i2},\mathit{n}) |_3= 1}\)
\end{program}  
where  \(\mathtt{|\mathit{L}|_i = 0}\), if $L$ is not defined in $R_{i}$.  
  
The program ${\tt DIFF} := R_{1} \cup R_{2}\cup R_{3}$ is well-moded.    
  Hence,
by Proposition   \ref{pro:strongbound_wm},     
every  well-moded query is strongly bounded wrt.\  ${\tt DIFF}$ and    
$|\ |_1$, $|\ |_2$, $|\ |_3$.      
\end{example}      
     
Note that the class of strongly bounded queries is generally     
larger than the class of well-moded queries.    
Consider for instance  the  program \texttt{MOVE} and the query      
$Q:=$ $\mathtt{move([X1,X2],Ys), delete(Ys,Y,Zs)}$   
which is not well-moded  since it is not ground in the input     
position of the first atom.    
However $Q$ can be easily recognized to be strongly    
bounded  wrt.\   \texttt{MOVE}  and   
 $|\ |_{1}$, $|\ |_{2}$   
defined in Example \ref{exa:well-moded-program1}.  
    We will come back to this query later.
  
\subsection{Well-Typed Programs}  
  
A more refined well-behavior property of programs,  
namely well-typedness, can also be useful in order to ensure the   
strong boundedness property. 
  
The notion of well-typedness relies both on the concepts of \emph{mode}     
and \emph{type}. The following very general definition of a type is     
sufficient for our purposes.

\begin{definition}[Type]      
A \emph{type} is a set of terms closed under substitution.       
\end{definition}      
      
Assume as given a specific set of types,   
denoted by \emph{Types}, which includes       
$Any$, the set of all terms, and $Ground$ the set of all ground terms.      
      
\begin{definition} [Type Associated with a Position]     
A \emph{type for an $n$-ary predicate symbol $p$} is      
a function $t_p$ from $\{1,\ldots,n\}$ to the set \emph{Types}.      
If $t_p(i)=T$, we call $T$ \emph{the type associated with the      
position $i$ of $p$}.      
Assuming a type $t_p$ for the predicate $p$, we say that a     
literal  $p(s_1,\ldots,s_n)$ is \emph{correctly typed in position $i$}      
if $s_i\in t_p(i)$.       
\end{definition}      
      
In a typed program we assume that every predicate $p$ has a fixed     
mode $m_p$ and a fixed type $t_p$ associated with it and we denote     
it by \[p(m_p(1):t_p(1),\ldots,m_p(n):t_p(n)).\]    
So, for instance, we write      

\[\mathtt{append(\mathit{+:List},\mathit{+:List},\mathit{-:List})}\]   
to denote the moded atom    
$\mathtt{append(\mathit{+},\mathit{+},\mathit{-})}$    
where the type associated with each argument position is    
$\mathit{List}$, i.e., the set of all lists.    
    
We can then talk about types of input and of output positions of an    
atom.

The notion of well-typed queries and programs relies on the   
following concept  of type judgement.  
  
\begin{definition}[Type Judgement]  
 By a \emph{type judgement} we mean a statement of the form  
\(\mathbf{s:S\Rightarrow t:T}.\)  
 We say that a type judgement \(\mathbf{s:S\Rightarrow t:T}\)  
\emph{is true}, and write   
\(\mathbf{\models s:S\Rightarrow t:T},\)  
if for all substitutions $\theta$,  
\(\mathbf{s}\theta\in \mathbf{S}\) implies   
\(\mathbf{t}\theta\in \mathbf{T}\).   
\end{definition}  
  
For example, the type judgements \((x: \mathit{Nat}, \; l:  
\mathit{ListNat}) \Rightarrow ([x|l]: \mathit{ListNat})\)  
and \(( [x|l]: \mathit{ListNat}) \Rightarrow (l:\mathit{ListNat})\)  
are both true.  
  
A notion of well-typed program has been first introduced in
 \cite{BLR92} and also studied  in 
\cite{AE93} and in \cite{AL95}.  Similarly to  
well-moding, the notion was developed for definite programs.  
Here we extend it to general programs.  
  
 In the following definition, we assume that  
 $\mathbf{i}_s: \mathbf{I}_s$  
is the sequence of typed terms filling in the input positions of $L_s$  
and $\mathbf{o}_s:\mathbf{O}_s$
  is the sequence of typed terms filling in the  
output positions of $L_s$.  
  
\begin{definition}[Well-Typed]  
\label{def:well-typed}  
\begin{itemize}  
\item  
A query \(L_1,\ldots, L_n\)  
is called \emph{well-typed}  
if for  all $j\in\{1,\ldots,n\}$   
\[\models \mathbf{o}_{j_1}: \mathbf{O}_{j_1}, \ldots, \mathbf{o}_{j_k}: \mathbf{O}_{j_k} \Rightarrow  
\mathbf{i}_{j}: \mathbf{I}_{j}\]  
where $L_{j_1},\ldots , L_{j_k}$ are all the positive literals 
in $L_1,\ldots,L_{j-1}$.  
\item   
A clause  \(L_0  
\leftarrow L_1,\ldots,L_n\)  
is called \emph{well-typed}  
if  for  all $j\in\{1,\ldots,n\}$ 
\[\models  \mathbf{i}_0: \mathbf{I}_0, \mathbf{o}_{j_1}: \mathbf{O}_{j_1}, \ldots, \mathbf{o}_{j_k}: \mathbf{O}_{j_k} \Rightarrow  
\mathbf{i}_j: \mathbf{I}_{j}\]  
where $L_{j_1},\ldots , L_{j_k}$ are all the  positive literals 
in $L_1,\ldots,L_{j-1}$, and  
\[\models  \mathbf{i}_0: \mathbf{I}_{0}, \mathbf{o}_{j_1}: \mathbf{O}_{j_1}, \ldots, \mathbf{o}_{j_h}: \mathbf{O}_{j_h} \Rightarrow  
\mathbf{o}_{0}: \mathbf{O}_{0}\]  
where $L_{j_1},\ldots , L_{j_h}$ are all the  positive literals 
in $L_1,\ldots,L_{n}$.  
\item   
A program is called  \emph{well-typed} if all of its clauses are   
well-typed.   
\end{itemize}  
\end{definition}  
Note that an atomic query is well-typed iff it is correctly  
typed in its input positions and a unit clause   
\(p(\mathbf{s:S},\mathbf{t:T})\leftarrow\) is well-typed if \(\mathbf{\models  
s:S \Rightarrow t:T}\).  

The difference between Definition \ref{def:well-typed} and the one usually given   
for definite programs is that  the correctness  
of  the terms filling  
in the output positions of negative literals  
cannot be used to deduce  
the correctness of the terms filling  
in the input positions of a  literal to the right 
(or the output positions of the head in a clause).  
The two definitions coincide either for definite programs or  
for general programs whose negative literals 
have only input positions.
  
 As an example, let us consider the trivial program

\begin{program} 
\>  p($\mathit{-:List}$).\\
\>q($\mathit{+:List}$).\\[2mm]
\>p([]).\\
\> q([]).
\end{program}
By adopting a straightforward extension of well-typedness 
to normal programs which  considers also the outputs of negative literals,
we would have that the query $\mathtt{\neg  p(a), q(a)}$ is well-typed
even if $\mathtt{a}$ is not a list. Moreover  well-typedness would 
not be persistent wrt.\  LDNF-resolution since
$\mathtt{q(a)}$, which is the first LDNF-resolvent of the previous query, 
is no more well-typed.
Our extended definition and the classical one 
coincide either for definite programs or for 
general programs whose  negative literals have only input positions.
  
For definite programs, well-modedness can be viewed as a special case of
well-typedness if we consider only one type: $Ground$.
With our extended definitions of well-moded and well-typed 
general programs this is no more true.
We could have given a more complicated definition
for well-typedness in order to capture
 also well-modedness as a special case.
For the sake of simplicity, we prefer to give 
two distinct and simpler definitions.

\begin{remark}  
\label{remark:prefix_wt}  
If $Q$ is a well-typed query, then all its non-empty prefixes are well-typed.  
In particular, $\mathit{first}(Q)$ is well-typed.  
\end{remark}  
  
The following Lemma  
shows that well-typed queries are closed under LDNF-resolution.  
It has been proved in \cite{BLR92}  
for  definite programs.  
  
\begin{lemma}  
\label{lemma:well-typed-closed}  
Let $P$ and $Q$ be a well-typed program and query, respectively.
 Then all  
LDNF-descendants of $P\cup \{Q\}$ are well-typed.   
\end{lemma}  
\begin{proof}  
Similarly to the case of well-moded programs,  
to extend the result to general programs   
it is sufficient to  show that if a query $Q:=\neg A, L_1,\ldots,L_n$   
is  well-typed  
 then both $A$ and $L_1,\ldots,L_n$ are well-typed.  
In fact, by Remark \ref{remark:prefix_wt},  
$\neg A= \mathit{first}(Q)$ is well-typed   
and by Definition \ref{def:well-typed}, if the first literal in a well-typed query is  
negative, then it is not used  
to deduce  well-typedness of the rest of the query.  
\end{proof}

It is now natural to exploit well-typedness in order to check strong  
boundedness.  
Analogously to well-moded programs, there are level mappings that are  
more natural in presence of type information. They are the level mappings  
for which every well-typed atom is bounded.  
By Lemma \ref{lemma:well-typed-closed} we have that  
a well-typed query $Q$ is strongly bounded wrt.\  
a well-typed program $P$ and any such level mapping.   
This is stated by the next proposition.  
  
\begin{proposition}      
\label{pro:strongbound_wt}      
Let   $P:=R_1\cup \ldots \cup R_n$ be a \emph{well-typed} program   and 
 $ R_n\sqsupset \ldots \sqsupset R_1$ be a hierarchy,  and       
$|\ |_1,\ldots,|\ |_n$   be       
 level mappings for $R_1,\ldots,R_n$,      
respectively. Suppose that  for      
every well-typed atom $A$, if $A$ is defined in $R_i$       
then $A$ is bounded wrt.\   $|\ |_i$,      
for $i\in\{1,\ldots,n\}$.         
 Then every well-typed query is      
strongly bounded wrt.\  $P$ and $|\ |_1,\ldots,|\ |_n$.      
\end{proposition}

\begin{example}      
\label{exa:typed1}    
Let us consider again the modular proof of      
termination for  $\mathtt{MOVE}:=R_{1}\cup R_{2}$, where        
$R_{1}$ defines  the relations       
\texttt{append1}, \texttt{append2} and \texttt{delete},      
while $R_{2}$, which extends $R_{1}$,  defines the relation       
\texttt{move}.      
We consider the moding of  Example~\ref{exa:well-moded-program1} with the       
following types:      
\begin{program}      
\>  \(\mathtt{delete(\mathit{+:List},\mathit{-: Any},\mathit{-:List})}\)\\      
\>  \(\mathtt{append1(\mathit{-:List},\mathit{-:List},\mathit{+:List}}\)) \\      
\>  \(\mathtt{append2(\mathit{+:List},\mathit{+:List},\mathit{-:List})}\)\\      
\>  \(\mathtt{move(\mathit{+:List},\mathit{-:List})}\).      
\end{program}          
 Program $\mathtt{MOVE}$ is \emph{well-typed} in the assumed modes and      
 types.      
      
 Let us consider the same      
level mappings as   
used in  Example~\ref{exa:well-moded-program1}.      
We have already seen that      
$R_{2}$  is acceptable wrt.\    $|\ |_{2}$ and any model, and    
$R_{1}$ is    acceptable wrt.\       
$|\ |_{1}$ and  any model.      
By definition of $|\ |_{2}$ and $|\ |_{1}$, one can easily see that       
\begin{itemize}      
\item      
 every well-typed atom $A$ defined in $R_{i}$      
is bounded wrt.\   $|\ |_{i}$.      
\end{itemize}      
Hence, by Proposition \ref{pro:strongbound_wt},      
\begin{itemize}      
\item      
 every well-typed query   is strongly bounded wrt.\    
 $\mathtt{MOVE}$ and   $|\ |_{1}$, $|\ |_{2}$.      
\end{itemize}    
Let us consider again the query   
$Q:=\mathtt{move([X1,X2],Ys), delete(Ys,Y,Zs)}$ which is not
well-moded but it is well-typed.  
We have that $Q$ is strongly bounded wrt.\ {\tt MOVE}   
and  $|\ |_{1}$, $|\ |_{2}$,    
and consequently, by Theorem \ref{theo:strongn},    
that  all  LDNF-derivations  of       
${\tt MOVE} \cup\{Q\}$ are finite.     
\end{example}

\begin{example}      
\label{exa:typed2}      
Consider the program \texttt{COLOR\_MAP} from      
\cite{SS86} which gene\-rates a coloring of a map in such       
a way that no two neighbors have the same color.      
The map is represented as a list of regions and colors as a list of       
available colors. In turn, each region is determined by its name, 
color and the      
colors of its neighbors, so it is represented as a term      
\texttt{region(name,color,neighbors)}, where \texttt{neighbors}      
is a list of colors of the neighboring regions.      
      
\begin{program2}      
\> c1: \> color\_map([ ],Colors).\\      
\> c2: \> color\_map([Region|Regions],Colors) \la \\      
\> \> \> color\_region(Region,Colors),\\      
\> \> \> color\_map(Regions,Colors).\\[2mm]      
\> c3: \> color\_region(region(Name,Color,Neighbors),Colors) \la \\      
\> \> \> select(Color,Colors,Colors1)\\      
\> \> \> subset(Neighbors,Colors1).\\[2mm]      
\> c4: \> select(X,[X|Xs],Xs).\\      
\> c5: \> select(X,[Y|Xs],[Y|Zs]) \la  select(X,Xs,Zs).\\[2mm]      
\> c6: \> subset([ ],Ys). \\      
\> c7: \> subset([X|Xs],Ys) \la       
 member(X,Ys), subset(Xs,Ys).\\[2mm]      
\> c8: \> member(X,[X|Xs]).\\      
\> c9: \> member(X,[Y|Xs]) \la member(X,Xs).      
\end{program2}      
       
Consider the following modes and types for the program       
\texttt{COLOR\_MAP}:           
\begin{program}      
\>   \(\mathtt{color\_map}(\mathit{+: ListRegion}, \mathit{+: List})\)\\      
\>   \(\mathtt{color\_region}(\mathit{+: Region}, \mathit{+: List})\)\\      
\>   \(\mathtt{select}(\mathit{+:  Any}, \mathit{+: List}, \mathit{-: List})\)\\      
\>   \(\mathtt{subset}(\mathit{+: List}, \mathit{+: List})\)\\      
\>   \(\mathtt{member}(\mathit{+: Any}, \mathit{+: List})\)      
\end{program}      
  where    
\begin{itemize}      
\item \emph{Region} is the set of all terms of the form      
\texttt{region(name,color,neighbors)} with       
\(\mathtt{name},\mathtt{color}\in \mathit{Any}\) and      
\(\mathtt{neighbors}\in  \mathit{List}\),      
\item \emph{ListRegion} is the set of all lists of regions.      
\end{itemize}

We can  check that \texttt{COLOR\_MAP} is well-typed        
in the assumed modes and types.

We can divide the program \texttt{COLOR\_MAP} into four distinct modules,      
$R_1,R_2,R_3,R_4$, in the hierarchy     
$ R_4\sqsupset R_3 \sqsupset R_2 \sqsupset R_1$ as follows:      
\begin{itemize}      
\item $R_4:=\{\mathtt{c1}, \mathtt{c2}\}$ defines the relation      
\texttt{color\_map},      
\item $R_3:=\{\mathtt{c3}\}$ defines the relation      
\texttt{color\_region},      
\item $R_2:=\{\mathtt{c4}, \mathtt{c5},\mathtt{c6}, \mathtt{c7}\}$      
defines the relations       
\texttt{select} and \texttt{subset},      
\item $R_1:=\{\mathtt{c8}, \mathtt{c9}\}$        
defines the relation      
\texttt{member}.      
\end{itemize}       
      
Each $R_i$ is trivially acceptable wrt.\       
any model $M$ and  the simple level mapping $|\ |_i$  defined below:       
       
\begin{program}      
\> \(\mathtt{|color\_map(\mathit{xs},\mathit{y}s)|_{4} = |\mathit{xs}|_{length}}\)\\[2mm]      
\> \(\mathtt{|color\_region(\mathit{x},\mathit{xs})|_{3} = 1}\)\\[2mm]      
\> \(\mathtt{|select(\mathit{x},\mathit{xs},\mathit{ys})|_{2} = |\mathit{xs}|_{length}}\)\\      
\> \(\mathtt{|subset(\mathit{xs},\mathit{ys})|_{2} = |\mathit{xs}|_{length}}\)\\[2mm]      
\> \(\mathtt{|member(\mathit{x},\mathit{xs})|_{1} = |\mathit{xs}|_{length}}\)      
\end{program}      
where for all $i\in\{1,2,3,4\}$,      
$|L|_{i} = 0$, if $L$ is not defined in $R_i$.       
      
Moreover,    for  
 every well-typed atom $A$ and  $i\in\{1,2,3,4\}$,  
 if $A$ is defined in $R_i$ then $A$  
is bounded wrt.\    $|\ |_i$.  
Hence, by Proposition \ref{pro:strongbound_wt},      
\begin{itemize}      
\item      
every well-typed query   is strongly bounded      
wrt.\  the program $\mathtt{COLOR\_MAP}$ and      
$|\ |_1,\ldots,|\ |_4$.      
\end{itemize}      
This proves that      
all  LDNF-derivations  of the program \texttt{COLOR\_MAP}      
star\-ting in a well-typed query are finite.      
 In particular, all the  LDNF-derivations starting in a query      
of the form \(\mathtt{color\_map(\mathit{xs},\mathit{ys})} \), where      
$\mathit{xs}$  is a list of regions and       
$\mathit{ys}$  is a list,      
are finite.      
Note that in proving termination of
such queries the choice of a model is irrelevant.
Moreover, since such queries are well-typed,
their input arguments are required
to have a specified structure, but they
are not required to be ground terms as in the
case of well-moded queries.
Hence, well-typedness  allows us
to reason about a larger class of queries with
respect to well-modedness.
      
This example is also discussed  in \cite{AP94}.      
In order to prove its termination      
they define a particular level mapping $|\ |$, obtained by combining the level mappings  of each module, and a special model $M$      
wrt.\  which the whole program \texttt{COLOR\_MAP} is acceptable.      
Both the level mapping  $|\ |$ and the model $M$ are non-trivial.        
\end{example}

\section{Iterative Proof Method}      
\label{sec:iteration}  
    
In the previous section we have seen how we can exploit     
properties which are preserved by LDNF-resolution,    
such as well-modedness and well-typedness, for developing a    
modular proof of termination in a hierarchy of programs.    
In this section we show how these properties allow us to   
apply our general result, i.e., Theorem \ref{theo:generale-sulle-classi},  
also in an iterative way.

\begin{corollary}      
    \label{theo:R-well-terminating}      
   Let $P$ and $R$ be two  programs such that    
$P\cup R$ is well-moded and $P$ extends      
  $R$, and $M$ be a complete model of $P\cup R$.      
Suppose that      
\begin{itemize}      
\item $P$ is acceptable wrt.\  a moded level mapping $|\ |$ and $M$,      
\item for all well-moded queries $Q$,   
all  LDNF-derivations $R\cup\{Q\}$ are finite.  
\end{itemize}      
Then   for all well-moded queries $Q$,  
all  LDNF-derivations of  
$(P\cup R)\cup \{Q\}$ are finite.  
\end{corollary}      
\begin{proof}    
Let ${\cal C}$ be the class of well-moded queries of $P\cup R$.  
By Remark \ref{remark:prefix_wm}  
and  
Lemma \ref{lemma:well-moded-closed},  
${\cal C}$ is $(P\cup R)$-closed.  
Moreover  
\begin{itemize}  
    \item   
    $P$ is acceptable wrt.\  a moded level mapping $|\ |$ and $M$,    
    by hypothesis;  
\item   
for all well-moded queries $Q$, all LDNF-derivations of $R\cup\{Q\}$  
are finite, by hypothesis;  
\item   
for all  well-moded atoms $A$,  
if $A$ is defined in $P$ then $A$ is bounded wrt.\  $|\ |$,   
by Remark \ref{remark-modedlevelmapp},  since $|\ |$ is a moded  
level mapping.  
\end{itemize}  
Hence by Theorem \ref{theo:generale-sulle-classi} we get the thesis.  
\end{proof}    
\medskip      
    
Note that this result allows one to incrementally prove     
well-termination for general programs  thus   
extending the result given  in \cite{EBC99}
for definite programs.

A similar result can be stated also for well-typed programs and     
queries, provided that there exists a  level mapping  for $P$
implying    
boundedness of atomic well-typed queries.    
    
\begin{corollary}      
    \label{theo:R-typed-terminating}      
    Let $P$ and $R$ be two programs such that $P\cup R$ is    
 well-typed and $P$ extends      
    $R$, and $M$ be a complete model of $P\cup R$.  Suppose that      
\begin{itemize}      
\item $P$ is acceptable wrt.\  a level mapping $|\ |$ and $M$,    
\item every well-typed atom defined in $P$ is bounded wrt.\   $|\ |$,    
\item  for all well-typed queries $Q$,  
all  LDNF-derivations of  
$R\cup\{Q\}$ are finite. 
\end{itemize}      
Then   for all well-typed queries $Q$,  
all  LDNF-derivations of  
$(P\cup R)\cup \{Q\}$ are finite.  
\end{corollary}     
  
\begin{proof}    
Let ${\cal C}$ be the class of well-typed queries of $P\cup R$.  
By Remark \ref{remark:prefix_wt}  
and  
Lemma \ref{lemma:well-typed-closed},  
${\cal C}$ is $(P\cup R)$-closed.  
Moreover  
\begin{itemize}  
    \item   
    $P$ is acceptable wrt.\  a level mapping $|\ |$ and $M$,    
    by hypothesis;  
\item   
for all well-typed queries $Q$, all LDNF-derivations of $R\cup\{Q\}$  
are finite,  
by hypothesis;  
\item   
for all  well-typed atoms $A$,  
if $A$ is defined in $P$ then $A$ is bounded wrt.\  $|\ |$,  
by hypothesis.  
\end{itemize}  
Hence by Theorem \ref{theo:generale-sulle-classi} we have the thesis.  
\end{proof}

\begin{example}    
\label{exa:iterative-method1}  
Let us consider again the program \texttt{COLOR\_MAP}       
with the same modes and types     
as in Example \ref{exa:typed2}.    
We apply the iterative termination proof    
given by Corollary~\ref{theo:R-typed-terminating}    
to \texttt{COLOR\_MAP}.  
  
{\bf First step.}      
\indent  
We can consider at first two trivial modules,      
$R_{1}:=\{\mathtt{c8}, \mathtt{c9}\}$      
which defines the relation \texttt{member},      
and $R_{0}:=\emptyset $.  
We already know that     
\begin{itemize}    
    \item    
    $R_{1}$ is  acceptable wrt.\   any model $M$ and      
    the level mapping $|\ |_1$  already defined;    
\item     
all well-typed atoms $A$, defined in $R_{1}$,   
are bounded wrt.\    $|\ |_1$;    
\item   for all well-typed queries $Q$,  
all  LDNF-derivations of  
$R_{0}\cup\{Q\}$ are trivially finite.  
\end{itemize}    
Hence, by Corollary \ref{theo:R-typed-terminating},    
for all well-typed queries $Q$,  
all  LDNF-derivations of  
$(R_{1}\cup R_{0})\cup\{Q\}$ are finite.  
  
{\bf Second step.}  
\indent  
We can now iterate the process one level up.  
Let us consider the two modules,      
$R_{2}:=\{\mathtt{c4}, \mathtt{c5},\mathtt{c6}, \mathtt{c7}\}$      
which defines the relations       
\texttt{select} and \texttt{subset},      
and $R_{1}:=\{\mathtt{c8}, \mathtt{c9}\}$        
which defines the relation      
\texttt{member} and it is equal to $(R_{1}\cup R_{0})$   
of the previous step.      
We already showed in Example \ref{exa:typed2} that     
\begin{itemize}    
    \item    
    $R_{2}$ is  acceptable wrt.\   any model $M$ and      
    the level mapping $|\ |_2$  already defined;    
\item     
 all well-typed atoms $A$, defined in $R_{2}$,   
 are bounded wrt.\    $|\ |_2$;      
 \item   for all well-typed queries $Q$,  
all  LDNF-derivations of  
$R_{1}\cup\{Q\}$ are finite.  
\end{itemize}    
Hence, by Corollary \ref{theo:R-typed-terminating},    
 for all well-typed queries $Q$,  
all  LDNF-derivations of  
$(R_{2}\cup R_{1})\cup\{Q\}$ are finite.  
  
By iterating the same reasoning for two steps more,
we can prove  that all  LDNF-derivations      
of the program \texttt{COLOR\_MAP}      
starting in a well-typed query  are finite.      
\end{example}
Our iterative method applies     
to a hierarchy of programs    
where on the lowest module, $R$,     
we require termination wrt.\  a particular  class of queries.    
This can be a weaker requirement on $R$ 
than acceptability as shown in the following contrived example.  
\begin{example}    
\label{exa:iterative-method2}  
Let $R$  define the     
predicate \texttt{lcount} which counts     
the number of natural numbers in a list.    
    
\begin{program2}      
\>  \(\mathtt{lcount(\mathit{+:List},\mathit{-:Nat})}\)\\      
\>  \(\mathtt{nat(\mathit{+:Any})}\). \\[2mm]  
\> r1: \> lcount([ ],0).\\      
\> r2: \> lcount([X|Xs],s(N)) \la nat(X), lcount(Xs,N).\\      
\> r3: \> lcount([X|Xs],N) \la $\neg$ nat(X), lcount(Xs,N).\\    
\> r4: \> lcount(0,N) \la  lcount(0,s(N)).\\[2mm]    
\> r5: \> nat(0).\\    
\> r6: \> nat(s(N)) \la nat(N).    
\end{program2}

$R$ is well-typed wrt.\ the specified modes and types.   
Note that $R$ cannot be acceptable   
due to the presence of clause \texttt{r4}.   
On the other hand, the program    
terminates for all well-typed queries.    
    
Consider now the following program $P$ which extends $R$.  
The predicate {\tt split}, given a list of lists, separates the list  
elements containing more than {\tt max} natural numbers from the other   
lists:    
\begin{program2}      
\>   
\( \mathtt{split(\mathit{+:ListList},\mathit{-:ListList},\mathit{-:ListList})}\) \\  
\>  >\((\mathit{+:Nat},\mathit{+:Nat})\) \\  
\>  <=\((\mathit{+:Nat},\mathit{+:Nat})\) \\[2mm] 
\> p1: \> split([ ],[ ],[ ]).\\      
\> p2: \> split([L|Ls],[L|L1],L2) \la lcount(L,N), N > max, \\
\> \> \> split(Ls,L1,L2).\\    
\> p3: \> split([L|Ls],L1,[L|L2]) \la lcount(L,N), N <= max, \\
\> \> \> 
split(Ls,L1,L2).    
\end{program2}

where \textit{ListList}  
 denotes the set of all lists of lists, and \texttt{max} is    
a natural number. The program $P\cup R$ is well-typed.     
Let us consider    
the simple level mapping $|\ |$ for $P$ defined by:      
      
\begin{program}      
\> \(\mathtt{|split(\mathit{ls},\mathit{l1},\mathit{l2})| = |\mathit{ls}|_{length}}\)   
\end{program}      
   which assigns level $0$ to any literal not defined in $P$.     
Note that    
\begin{itemize}      
\item $P$ is acceptable wrt.\  the  level mapping $|\ |$ and any complete    
model $M$,      
\item all well-typed atoms defined in $P$ are bounded wrt.\   $|\ |$,   
\item for all well-typed queries $Q$,  
all  LDNF-derivations of  
$R\cup\{Q\}$ are finite.  
\end{itemize}    
Hence, by Corollary \ref{theo:R-typed-terminating},   
 for all well-typed queries $Q$,  
all  LDNF-derivations of  
$(P\cup R)\cup\{Q\}$ are finite.  

This example shows that well-typedness 
could be useful to exclude  what
might  be called ``dead code''.
 \end{example}

\section{Comparing with Apt and Pedreschi's Approach}  
\label{sec:comparisons}      
   
Our work can be seen as an extension of  a proposal in \cite{AP94}.  
Hence we devote this section to a comparison  
with their approach.   
    
On one hand, since our approach applies to general programs, it clearly covers   
cases which cannot be treated with the method proposed in \cite{AP94},  
which was developed for definite programs.  
On the other hand, for definite programs the classes of queries and programs   
which can  
be treated by Apt and Pedreschi's approach 
are properly included in those which can be treated by
our method as we show in this section. 
  
We first recall
 the notions of \emph{semi-acceptability} and \emph{bounded query}  
used in~\cite{AP94}.

\begin{definition}[Semi-acceptable Program]      
 Let $P$ be a definite program, $|\ |$ be a level mapping for $P$ and $M$ be
a       model of $P$.      
 $P$ is called \emph{semi-acceptable wrt.\  $|\ |$ and $M$}  if for every      
 clause $A\leftarrow {\bf A}, B,{\bf B}$ in ${\it ground}(P)$ such that     
 $ M\models {\bf A}$   
 \begin{itemize}    
   \item $|A|>|B|, \mbox{ if } rel(A) \simeq rel(B)$,  
   \item $ |A|\geq |B|, \mbox{ if } rel(A) \sqsupset rel(B)$.  
 \end{itemize}  
\end{definition}    


\begin{definition}[Bounded Query]
\label{def-ext-Boundedness}
Let $P$ be a  definite program, $|\ |$ be a level mapping for $P$, and
$M$ be a  model of $P$.
\begin{itemize}
\item[$\bullet$] With each  query $Q:=L_1,\ldots,L_n$
we associate $n$ sets of natural numbers defined as follows:
For $i\in \{1,\ldots,n\}$,
$$|Q|_i^M=\{|L'_i|\;|\; L'_1,\ldots,L'_n 
\mbox{ is a ground instance of } Q
\mbox{ and } M\models  L'_1,\ldots,L'_{i-1}\}.$$
\item[$\bullet$] A query $Q$ is called \emph{bounded wrt.
$|\ |$ and $M$} if $|Q|_i^M$ is finite
(i.e.,  if $|Q|_i^M$ has a maximum in ${\bf N}$)  for all $i\in\{1,\ldots,n\}$.
\end{itemize}
\end{definition}
  
\begin{lemma}  
\label{lemma:semi-preservation}  
Let  $P$ be a definite program  
which is semi-acceptable wrt.\  $|\ |$ and $M$.  
If $Q$ is a query bounded wrt.\    $|\ |$ and $M$  
then all  LD-descendants of $P\cup\{Q\}$ are bounded  
wrt.\   $|\ |$ and $M$.  
\end{lemma}  
\begin{proof}  
It is a consequence of Lemma 3.6 in \cite{AP94}
and (the proof of) Lemma 5.4 in \cite{AP94}.  
\end{proof}  
\medskip

We can always decompose a definite program $P$ into a hierarchy of   
$n \geq 1$  programs $P:=R_1\cup \ldots \cup R_n$, where  
$ R_n\sqsupset \ldots \sqsupset R_1$ in such a way that for every   
$ i \in \{1, \ldots, n\}$ if the predicate symbols   
$p_i$ and $q_i$ are both defined in $R_i$  
then neither $p_i \sqsupset q_i$ nor $q_i \sqsupset p_i$
(either they are mutually recursive or independent). 
 We call such a hierarchy a   
\emph{finest decomposition}   
of $P$. 

The following property has two main applications.
First it allows us to compare our approach with \cite{AP94},
then it provides  an extension of Theorem \ref{theo:strongn} 
to hierarchies of semi-acceptable programs.
  
\begin{proposition}  
    \label{prop:semi}  
Let $P$ be a semi-acceptable program wrt.\  a level mapping $|\ |$   
and a model $M$ and
$Q$ be a query strongly bounded wrt. $P$ and $|\ |$.
 Let $P := R_1 \cup \ldots \cup R_n$   
be a finest decomposition of $P$ into a hierarchy of modules.   
Let $|\ |_{i}$, with $i\in\{1,\ldots,n\}$, be defined in the following way:    
if $A$ is defined in $R_i$ then  
 $ |A |_{i}=|A|$ else $ |A |_{i}=0$.  
Then 
\begin{itemize}
\item every $R_i$
is acceptable wrt.\  $|\ |_{i}$ and $M$ (with $i\in \{1,\ldots,n\}$),
\item $Q$ is strongly bounded wrt. $R_1 \cup \ldots \cup R_n$
and $|\ |_{1},\ldots,|\ |_{n}$.
\end{itemize}
\end{proposition}  
\begin{proof}  
 Immediate by the definitions of semi-acceptability and 
strongly boundedness, since 
we are considering a finest decomposition.  
\end{proof}  
\medskip  
    
In order to compare our approach to the one presented in \cite{AP94}    
we consider only Theorem 5.8 in \cite{AP94}, since    
this is their most general result which implies  the other ones, namely  
Theorem 5.6 and Theorem~5.7.

\begin{theorem} [Theorem 5.8 in \cite{AP94}]    
\label{theo:apt-modular}    
Let $P$ and $R$ be two definite programs such that $P$ extends $R$, and let $M$ be    
a model of $P\cup R$.  Suppose that  
\begin{itemize}    
\item $R$ is semi-acceptable wrt.\  $|\ |_R$ and $M\cap B_R$,    
\item $P$ is semi-acceptable wrt.\  $|\ |_P$ and $M$,    
\item there exists a level mapping $|\!|\ |\!|_P$ such that for every    
ground instance  of a clause from $P$, $A\la \ol A, B, \ol B$, such that    
$M\models \ol A$    
\begin{itemize}    
\item  $|\!|A |\!|_P \geq |\!|B |\!|_P$, if $\mathit{rel}(B)$ is defined in $P$,    
\item $|\!|A |\!|_P \geq |B |_R$, if $\mathit{rel}(B)$ is defined in $R$.    
\end{itemize}    
\end{itemize}     
Then $P\cup R$ is semi-acceptable wrt.\  $|\ |$  and $M$, where $|\ |$    
is defined  as follows:    
\begin{itemize}      
\item[] $|A|=|A|_P + |\!|A |\!|_P$, if  $\mathit{rel}(A)$ is defined in $P$,    
\item[] $|A|=|A|_R$, if  $\mathit{rel}(A)$ is defined in $R$.    
\end{itemize}      
\end{theorem}

The following remark follows from Lemma 5.4 in \cite{AP94} and 
Corollary 3.7  in \cite{AP94}. Together with Theorem \ref{theo:apt-modular},
it implies termination of bounded queries in~\cite{AP94}.

\begin{remark}
If  $P\cup R$ is semi-acceptable wrt.\  $|\ |$  and $M$ and $Q$ is bounded
 wrt.\  $|\ |$  and $M$ then all LD-derivations of $(P\cup R)\cup\{Q\}$
are finite.
\end{remark}

We now  show that  
whenever  Theorem \ref{theo:apt-modular}
can be applied to prove termination  
of all the queries bounded wrt.\  $|\ |$ and $M$,  
then also our method can be used to  
prove termination of the same class of queries
with no need of  $|\!|\ |\!|_P$ for relating the proofs of the 
two modules.

In the following theorem for the sake of simplicity   
we assume that  $P \sqsupset R$ is a finest   
decomposition of  $P \cup R$. We 
discuss later how to
 extend the  result   to the general case.

\begin{theorem}    
Let $P$ and $R$ be two programs such that $P$ extends $R$, and let $M$ be    
a model of $P\cup R$.  Suppose that  
\begin{itemize}    
\item $R$ is semi-acceptable wrt.\  $|\ |_R$ and $M\cap B_R$,    
\item $P$ is semi-acceptable wrt.\  $|\ |_P$ and $M$,
\item there exists a level mapping $|\!|\ |\!|_P$ defined as in Theorem   
\ref{theo:apt-modular}. 
\end{itemize}  
Let $|\ |$ be the level mapping defined by Theorem   
\ref{theo:apt-modular}. Moreover, suppose $P\sqsupset R$ is  
a finest decomposition of $P\cup R$.  
If $Q$ is bounded wrt.\  $|\ |$, then $Q$ is strongly bounded wrt.\    
$P \cup R$  and  $|\ |_P$ and $|\ |_R$.  
\end{theorem}  
\begin{proof}    
Since we are considering a finest decomposition of  
$P\cup R$,  by Proposition~\ref{prop:semi},  
$R$ is acceptable wrt.\  $|\ |_R$, while $P$ is  
acceptable wrt.\  $|\ |'_P$ such that  
 if $A$ is defined in $P$ then  
 $ |A |'_{P}=|A|_P$ else $ |A |'_{P}=0$.  
  
By Lemma \ref{lemma:semi-preservation}  
all LD-descendants of $(P\cup R)\cup \{Q\}$ are  bounded wrt.\     
$|\ |$ and $M$.    
By definition of boundedness,     
for all LD-descendants $Q'$ of $(P\cup R)    
\cup\{Q\}$,    ${\it first}(Q')$  is bounded wrt.\    $|\ |$.    
By definition  
of $|\ |$, for all atoms  $A$ bounded wrt.\  $|\ |$ we have that:  
if $A$ is defined in $R$ then $A$ is bounded wrt.\  $|\ |_R$,  
 while if  $A$ is defined in $P$ then $A$ is bounded wrt.\  $|\ |_P$  
and hence  wrt.\  $|\ |'_P$ (since  $|A |'_P =|A |_P$).  
Hence the thesis follows.  
\end{proof}    
\medskip  
    
  If the hierarchy $P \sqsupset  R$ is not   
a finest one and  
$|\ |_{P}$ and $|\ |_{R}$ are the level mappings  
corresponding to $P$ and $R$ respectively, 
then we can decompose $P$  
into a finest decomposition, $P:= P_n\sqsupset \ldots \sqsupset P_1$ ,  
and consider instead   
of $|\ |_{P}$  the derived level   
mappings $|\ |_{P_i}$  defined in the following way:    
if $A$ is defined in $P_i$ then  
$ |A |_{P_i}=|A|_{P}$ else $ |A |_{P_i}=0$.  
Similarly we can decompose $R:= R_n\sqsupset \ldots \sqsupset R_1$  
and define the corresponding level mappings.  
The derived level mappings  
 satisfy all the properties we need for proving that
if $Q$ is bounded wrt.\  $|\ |$, then $Q$ is strongly bounded wrt.\    
$P \cup R$  and  $|\ |_{P_1}, \ldots,|\ |_{P_n},
|\ |_{R_1}, \ldots,|\ |_{R_n} $.\\

To complete the comparison with \cite{AP94},  
we can observe that  
our method is applicable  also for proving termination  
of queries in modular programs   
which are not (semi-)acceptable.  
Such programs clearly cannot be  
dealt with Apt and Pedreschi's method.  
The  program of Example \ref{exa:iterative-method2}  
is a non-acceptable program for which we proved   
termination of all well-typed queries by applying   
Corollary \ref{theo:R-typed-terminating}.  
The following is a simple example of a non-acceptable  
program to which we can  apply  
the  general Theorem \ref{theo:strongn}.

\begin{example}    
\label{exa:more-programs}    
   
 Let $R$ be the following trivial program:    
    
\begin{program2}      
\> r1: \> q(0).\\      
\> r2: \> q(s(Y)) \la q(Y).    
\end{program2}      
  
The program $R$ is acceptable wrt.\  the following    
natural level mapping $|\ |_R$    and any model $M$:    
      
\begin{program}      
\> \(\mathtt{|q(\mathit{t})|}_R = \mathtt{|\mathit{t}|}_{size}\).    
\end{program}      
  
Let $P$ be a program, which extends $R$, defined as follows:    
    
\begin{program2}      
\> p1: \> r(0,0).\\      
\> p2: \> r(s(X),Y).\\      
\> p3: \> p(X) \la r(X,Y), q(Y).    
\end{program2}      
    
The program $P$ is acceptable wrt.\  the following  
trivial level mapping $|\ |_P$    and any model $M$:    
      
\begin{program}      
\> \(\mathtt{|q(\mathit{y})|}_P = 0\),\\    
\> \(\mathtt{|r(\mathit{x},\mathit{y})|}_P = 0\),\\    
\> \(\mathtt{|p(\mathit{x})|}_P = 1\).    
\end{program}      
    
Note that, even if each module is acceptable,
 $P\cup R$ cannot be acceptable wrt.\  any level mapping    
and model. In fact  $P\cup R $ is not left-terminating: for example    
it does not terminate for the ground query \texttt{p(s(0))}.    
As a consequence Apt and Pedreschi's method    
does not apply to  $P\cup R$.    
On the other hand, there are ground queries,  
such as  \texttt{p(0)},  which terminate in $P\cup R$.  
We can prove it as follows.

\begin{itemize}  
\item   
By  Theorem \ref{theo:strongn},  
for all strongly bounded queries $Q$ wrt.\  $P \cup R$ and   
$|\ |_R$, $|\ |_P$,  
 all LD-derivations of $(P\cup R)\cup\{Q\}$ are finite.  
\item   
\texttt{p(0)} is strongly bounded   wrt.\  $P \cup R$ and   
$|\ |_R$, $|\ |_P$.  
In fact, $\mathit{Call}_{(P\cup R)}(\texttt{p(0)})  
=\{\texttt{p(0)}, \texttt{r(0,Y)}, \texttt{q(0)}\}$   
and all these atoms are bounded wrt.\  their corresponding  level mapping.  
\end{itemize}  
\end{example}

\section{Conclusions}    
\label{conclusion}    

In this paper we propose a modular approach     
to termination proofs of general programs    
by  following the proof style introduced by Apt and Pedreschi.    
Our technique allows one to give simple proofs     
in hierarchically structured programs,     
namely programs which can be partitioned into $n$ modules,
$R_1\cup \ldots \cup R_n$,
 such that for all $i \in \{1,\ldots,n-1\}$,
$ R_{i+1}$ extends
$R_1\cup \ldots \cup R_{i}$.

We supply the general Theorem \ref{theo:generale-sulle-classi}  
which can be iteratively applied
to a hierarchy of two programs and a class of queries
enjoying persistence properties through LDNF-resolution.
We then use such a result to deal with a general hierarchy of
acceptable programs,  by introducing an extension of the     
concept of boundedness for hierarchical programs, namely     
strong boundedness.  
Strong boundedness is a property on queries which can be easily
ensured for hierarchies of  programs behaving well, such as 
well-moded or well-typed programs. We show how specific and simple
hierarchical termination proofs can be derived for such classes of
programs and queries.
We believe this is a valuable proof technique since realistic programs are 
typically  well-moded and well-typed.

The simplifications in the termination proof derive from the fact 
that for proving the termination of a modular program,
we simply prove acceptability of each module by 
choosing a level mapping which focuses
only on the predicates defined in it,
with no concern of the module context.
 Generally this can be done by using  very simple 
and natural level mappings which are completely independent
from one module to another.
A complicated level mapping is generally required when
we  prove  the termination of a  program as a whole
and  we have to consider a level mapping which appropriately     
relates all the predicates defined in the program. 
Hence the finer the modularization of the program    
the simpler the level mappings.  
Obviously we cannot completely ignore how predicates defined in
different modules relate to each other. 
On one hand, when we prove acceptability for each module,
we consider a model for the whole program.
This guarantees the compatibility among the definitions
in the hierarchy.
On the other hand, for queries we use the notion of strong boundedness.
The intuition is that we consider only what may influence the evaluation
of queries in the considered class.

The proof method of Theorem \ref{theo:generale-sulle-classi}
 can be applied also to programs which are not 
acceptable. In fact,  the condition on the
lower module is just that it terminates on all the queries in 
the considered class and not on all ground queries as required
for acceptable programs. From    Theorem \ref{theo:generale-sulle-classi}
 we could also derive a method to deal with 
pre-compiled modules (or even modules written in a different language)
provided that we already know termination properties
and we have a complete specification.

For sake of simplicity, in the first part of the paper we consider the 
notion of acceptability
instead of the  less requiring notion of
semi-acceptability. 
This choice makes proofs of our results much simpler.
On the other hand, as we show in Section \ref{sec:comparisons},
our results can be applied also to  hierarchies
of semi-acceptable programs.

We have compared our proposal  with
the one  in \cite{AP94}.
They propose a modular approach     
to left-termination proofs in a hierarchy of two definite programs 
$P \sqsupset R$.    
They require both the (semi)-acceptability of the two modules $R$ and $P$     
wrt.\  their respective level mappings and
a condition relating the two level mappings
which is meant to connect the two termination proofs.

Our method is more powerful both because we consider also general 
programs and because we capture  definite programs and 
queries which cannot be treated by the method developed in \cite{AP94}.
In fact there are
 non-acceptable programs for which we can single out
a class of terminating  queries.

For the previous reasons our method improves also with respect to 
\cite{PR96,PR99} where
hierarchies of modules are considered. In \cite{PR96,PR99} a 
unifying framework for the verification of total correctness 
of logic programs is provided. The authors consider modular termination 
by following the approach in \cite{AP94}. 

In \cite{Mar96} a methodology for proving termination of
general logic programs is proposed which is based on
modularization.
In this approach, the {\em acyclic} modules, namely 
modules that terminate independently from
the selection rule, play a distinctive 
role. For such modules, the termination proof
does not require a model. In combination with
appropriate notions of {\em up-acceptability} and {\em low-acceptability}
for the modules which are not acyclic,
this provides a practical technique for proving termination of the 
whole program.
Analogously to \cite{AP94},
also in \cite{Mar96} a relation between
the level mappings of  all modules is required.
It is interesting to note that the idea 
of exploiting acyclicity  is completely orthogonal
to our approach: we could integrate it into our framework.

Another related work is \cite{DDV99}, even if it does not 
aim explicitly at modularity.
In fact they propose a technique for automatic
termination analysis of definite programs which is highly efficient
also because they use a rather operational notion
of acceptability with respect to a set of queries,
where decreasing levels are required only on (mutually)
recursive calls as in \cite{DVB92}. Effectively, this 
corresponds to considering a finest decomposition of
the program and having independent level mappings
for each module. However, their notion of  acceptability
is defined and  verified  on call-patterns
instead of program clauses.
In a sense, such an acceptability with respect to a set
of queries combines the
concepts of strongly boundedness and (standard) acceptability.
They start from a class of queries and  try to  derive automatically a
termination proof  for such a class, while we start from
the program and   derive
a class of queries for which it terminates.

In \cite{VSD99} termination in the context of tabled execution is 
considered. Also in this case modular results are inspired by 
\cite{DVB92} 
by adapting the notion of acceptability wrt. call-patterns
to tabled executions. This work is further developed
in \cite{VSD01} where their modular termination conditions
are refined following the approach by \cite{AP94}.

In \cite{EBC99} a method for modular termination proofs for well-moded
definite programs is proposed.
 Our present work generalizes  such
result  to general programs.

Our method may help in designing more powerful automatic 
systems for veri\-fying termination \cite{DVB92,SSS97,DDV99,CT99}. 
We see two directions which could be pursued for a
fruitful integration with existing automatic tools. 
The first one exploits the fact that in each single module
it is sufficient to synthesize a level mapping which 
does not need to measure atoms defined in other modules.
The second one concerns tools based on call-patterns analysis 
\cite{DVB92,GG94,CD95}.
They can take advantage of the concept of strong boundedness which, as we
show, can be implied by well-behavior of programs \cite{DW88,Deb89}.

\paragraph{Acknowledgements.}  
This work has been partially supported by 
MURST with the National Research Project
``Certificazione automatica di programmi mediante interpretazione astratta''.

\bibliographystyle{dcu}      
\bibliography{BCER}      
  
\end{document}